# Design and characteristics of a WEP test in a sounding-rocket payload

Robert D. Reasenberg,[1] Biju R. Patla,[1] James D. Phillips,[1] and Rajesh Thapa[1,2]

[1] Smithsonian Astrophysical Observatory, Harvard-Smithsonian Center for Astrophysics, 60 Garden St., Cambridge, MA 02138.

[2] NP Photonics, UA Science & Technology Park, 9030 S. Rita Road, Suite 120, Tucson, AZ 85747

E-mail: reasenberg@cfa.harvard.edu

Abstract. We describe SR-POEM, a Galilean test of the weak equivalence principle that is to be conducted during the free fall portion of the flight of a sounding rocket payload. This test of a single pair of substances will have a measurement uncertainty of $\sigma(\eta) < 2 \times 10^{-17}$ after averaging the results of eight separate drops, each of 120 s duration. The entire payload is inverted between successive drops to cancel potential sources of systematic error. The weak equivalence principle measurement is made with a set of four of the SAO laser gauges, which have achieved an Allan deviation of 0.04 pm for an averaging time of 30 s. We discuss aspects of the current design with an emphasis on those that bear on the accuracy of the determination of $\eta$. The discovery of a violation ($\eta \neq 0$) would have profound implications for physics, astrophysics and cosmology.

PACS: 04.80

Keywords: WEP, EP, equivalence principle, general relativity, sounding rocket

## 1. Introduction

The weak equivalence principle (WEP) underlies general relativity, but violations are possible in most theories being developed to unify gravity with the other forces. A WEP violation is characterized by $\eta$, which is identically zero in any metric theory of gravity, including general relativity.

$$\eta_{AB} = \frac{a_A - a_B}{(a_A + a_B)/2} \qquad (1)$$

[e.g., Fischbach and Talmadge 1999] where $a_A$ and $a_B$ are the accelerations of bodies A and B, these bodies are moving under the influence of identical gravity fields, and there is no other cause of the acceleration. The discovery of a WEP violation would have profound implications for physics, astrophysics and cosmology.

The present best tests of the WEP are made using a rotating torsion pendulum and yield $\sigma(\eta) = 1.8 \times 10^{-13}$, consistent with $\eta = 0$. [Schlamminger et al. 2008]. Advancement of this approach has slowed because of intrinsic problems with the suspension fiber that may be overcome by operating at liquid helium temperature [Newman 2001, Berg et al. 2005]. Torsion balances at liquid helium temperature are also being used to study short range gravity [Hammond 2007]. There are several proposals for better WEP tests in an Earth-orbiting spacecraft [Nobili et al. 2009, Chhun et al. 2007]. The most sensitive among these is the satellite test of the equivalence principle (STEP) [Sumner 2007, Overduin et al. 2009].

STEP is a cryogenic experiment based on technological heritage from the Gravity Probe-B Mission [Everitt et al. 2011]. It is aimed at a measurement uncertainty of σ(η) < 10$^{-18}$.

## 2. Instrument Design and Operation

The sounding-rocket based principle of equivalence measurement (SR-POEM) will achieve a sensitivity σ(η) < 2 × 10$^{-17}$ in the ≈ 20 minute portion of a single flight of a suborbital rocket in which atmospheric drag is negligible. This is a five-fold improvement over our previous design [Reasenberg et al. 2011, Reasenberg and Phillips 2010]. We use subpicometer tracking frequency laser gauges (TFGs) to measure the distances from a highly stable "TFG plate," to two test masses (TMs), containing test substances "A" and "B". These data yield an estimate of the differential TM acceleration. We have presented the error budget for the previous design [Patla 2010]. A detailed error budget for the present version is in preparation and will be published soon [Patla et al. in preparation]. The measurements are made during eight "drops," each of duration $Q$ (= 120 s, see below). For a drop, the TMs initially have very nearly coincident centers of mass (CMs), and their common CM is near to that of the free falling payload. The TMs are nominally free during the WEP measurements, which is a distinguishing feature of SR-POEM. The eight drops, which are symmetrically distributed around apogee, alternate with inversions in which the entire payload is rotated 180 degrees to reverse the direction of the sensitive (vertical) axis. The inversions are implemented by the payload attitude control system (ACS), which fires cold-gas jets. In spacecraft coordinates, the inversions leave unchanged the systematic errors from most sources, but the WEP signal is reversed.

After the outer skin and nosecone are shed and the upper stage rocket motor has burned, it is separated from the payload in preparation for the ascent calibration phase. The payload comprises the physics package, the experiment support electronics and the payload support modules including power, telecommunications, navigation and ACS, and the mission sequencer (control computer). The physics package (Fig. 1) includes the dual vacuum chamber, everything within it and some appended devices. In particular, it includes the TMs, TM housing (capacitance gauge electrodes), TM caging hardware and all other objects that are near enough to the TMs to create significant non-uniform gravity gradients.

Gravity from local sources is kept constant at the TMs by the physics-package position servo (PPPS). The physics package is mounted on an active hexapod (Stewart platform), which is the thermally isolating (six degrees of freedom) actuator for the PPPS. During a drop, the PPPS moves the physics package (exclusive of the TMs, which are in free fall) to keep a constant orientation and spacing with respect to the freely falling TMs, thus holding constant to within servo error the contribution of the physics package to gravity at the TMs.

The mass of the physics package will be distributed symmetrically to within manufacturing tolerances. Residual asymmetries, thermal warping and shifts of the TM with respect to the physics package create spurious acceleration. Changes in this acceleration, synchronous with the inversion cycle, would cause systematic error. During the drops, the capacitance gauges will operate with a 3 mV rms drive signal (exclusive of the Z direction, which will have no drive signal). The capacitance gauge will measure changes of spacing between the TMs and physics package in the X and Y directions with a sensitivity of $6\ nm\ Hz^{-1/2}$. The Z-direction measurements are made by the TFGs with sensitivity of $0.1\ pm\ Hz^{-1/2}$ (see Section 2.2).

There are no moving parts such as reaction wheels. The ACS uses fiber-optic gyros. The ACS thrusters, which do have moving parts, are traditionally a source of small attitude and translation disturbance. We will take particular care with these by depressurizing their gas feed and disabling them prior to the drop. We will test parts on the ground, and if necessary provide extra shutoff valves to be closed during drops.

Let $\vec{r}$ and $\vec{\theta}$ be the average over a drop of the position and orientation of the physics package with respect to the average position and orientation of the two TMs. Both the PPPS error signal and the disturbances are consistent with holding $\vec{r}$ and $\vec{\theta}$ constant, both during a drop and from one drop to the next, to better than 1 $nm$ and $2 \times 10^{-8}$ $radian$.

The mass nearest to the TM is the TM housing. The 6 $kg$ housing lid alone would accelerate each TM upwards by $1.3 \times 10^{-8}$ $m/s^2$ ($2 \times 10^{-9} g(h_m)$). However, we expect to reduce this acceleration at least 30-fold by arranging the nearby mass with symmetry to minimize the force and force gradient (note thick TM housing top and walls in Figs. 1 and 2). We require the *variation* of the *difference* of Z-accelerations *in synchronism with the inversions* to be less than $10^{-17} m\ s^{-2}$ over the mission. Note that the TFGs measure TM positions with respect to the TFG plate, so the PPPS reduces the common Z acceleration that the TFGs measure and that would subsequently be subtracted in the analysis. Since the PPPS motion in Z is based on TFG measurements during the entire drop, the common acceleration is reduced to less than $3 \times 10^{-14}\ g(h_m)$, the acceleration that can be sensed with 3 $s$ of TFG data.

A "potato chip" bending of the TM housing lid would accelerate the TMs differentially. However, to exceed the above requirement, the amplitude would need to exceed 100 pm peak to zero. This deformation could be produced by thermal expansion if there were a top-bottom temperature gradient with azimuthal variation, but the temperature gradient required is of the order of 0.1 K, and this gradient would need to change in synchronism with the inversions to impact the estimate of η. The average temperature in the TM chamber is expected to be constant to within a few μK (see Sec. 2.10). The gradient will be kept small by a combination of the thermal conductivity of the vacuum chamber wall and by the symmetry of the chamber supports (Fig. 1). Thus, the differential TM acceleration due to such a thermal gradient is many orders of magnitude smaller than $10^{-17} m\ s^{-2}$.

Consider next a tilt of the housing lid. A tilt by an angle $\theta_x$ causes a difference of z-components of TM acceleration of

$$\Delta a_z = 4 \times 10^{-17} m/s^2 \left(\frac{\theta_x}{0.0001\ radian}\right)^2. \qquad (2)$$

With a tilt due to manufacturing error, a rotation of the physics package has a first-order effect on the differential acceleration of the TMs. Assuming an initial tilt of $0.0001\ radian$, the above error criterion is met if the change of angle from drop to drop is $< 10^{-5} radian$. The PPPS is expected to do better by almost three orders of magnitude. The tilted lid also causes the differential acceleration of the TMs to vary with their common x-position (i.e., with a variation of $r_x$):

$$\Delta \frac{\partial a_z}{\partial x} = 2 \times 10^{-13} s^{-2} \left(\frac{\theta_x}{0.0001\ radian}\right). \qquad (3)$$

If the tilt is $0.0001\ radian$, the x-component of the change of $\vec{r}$ need only be constant to within 50 μm. The PPPS is expected to do better by almost five orders of magnitude.

Suppose finally that the housing lid is displaced by 0.1 $mm$ in the X-direction (see Fig. 2). Then the change of differential Z-acceleration of the TMs due to movement of both TMs in the X-direction is $\Delta (da_z/dx) = 1 \times 10^{-11}\ s^{-2}$. The permitted change of $\vec{r}$ between drops is 1 μm. The PPPS is expected to do better by three orders of magnitude.

Thermal radiation from a 270°K Earth on a 500 kg spacecraft presenting a 1 $m^2$ reflective surface towards Earth accelerates the spacecraft with respect to an inertial frame by $3 \times 10^{-10}\ g(h)$, where $g(h_m) = 6.4\ ms^{-2}$ is the acceleration due to gravity at the mean altitude at which the experiment is performed, $h_m = 1500\ km$. The PPPS keeps the physics package stationary with respect to the TMs. Its initial conditions can be chosen so that the velocity of the physics package and TMs matches that of the spacecraft at the midpoint of the drop. Then, the change during the drop of the position of the physics package and TMs with respect to the spacecraft is 2 microns peak to zero. This is both smooth and mostly predictable.

By means of the control of the hexapod, the experiment attains some of the advantages of drag-free operation[1] and significant reduction is achieved in its rotation and translation with respect to the nearly inertial frame defined by the TMs. With the payload on a low-drag (sunlight free, high altitude and sub-orbital velocity) trajectory, the hexapod actuators do not need a large range of motion.

We can compare this experiment to a drag-free satellite; the test masses in both types of experiment are subject to many of the same forces. Consider the effect on the SR-POEM TMs before taking the difference of position (acceleration) between the two TMs and before cancellation due to inversion. We consider electrostatic attraction, TFG laser radiation pressure, gravitational attraction due to local sources, Coriolis pseudo-acceleration, pressure from residual gas, and magnetic force. The SR-POEM vacuum is $10^{-10}$ $Torr$ and there is a 2-layer magnetic shield that leaves only a small field gradient to interact with the intrinsic (remnant) magnetic moment of a TM. Of the sources considered, residual gas has the largest effect at $1 \times 10^{-12}$ $g(h_m)$. In Sec. 2.10, we discuss the stability of gas pressure and show that the gas-driven acceleration cancels due to inversion.

Overcoming the effects of gravity gradient is an easier problem for SR-POEM than for a drag-free satellite because, for SR-POEM, we are concerned with the *change* with inversion of the *difference* of accelerations of the two TMs. Further, since the TMs have nearly coincident CMs, a perturbing mass must interact with their (distinct) quadrupole moments to differentially accelerate them. This causes the perturbation of an asymmetrically placed point mass to fall off as $r^{-4}$ (cf. drag free satellite where the perturbation of a mass falls off as $r^{-2}$). In both cases, a change in the effect of a perturbing mass is proportional to an additional factor of $\Delta r/r$, where $\Delta r$ is the change of position of the mass (and angular dependences have been suppressed here). Finally, in SR-POEM, the CMs can be adjusted quickly: in a few seconds, the differential TM acceleration can be measured well enough to position the TM adequately for the WEP test. In a drag-free satellite, orbit determination must be performed and used in conjunction with the thruster history to determine the offset of the nominally centered TM from the payload CM.

Nonetheless, the SR-POEM requirement is severe. The PPPS provides additional suppression of this disturbance that is not (and would not be) used in drag-free satellites by keeping nearby perturbing masses from moving with respect to the TMs. Reducing gravity gradient disturbance requires an important calibration step. The quantity of interest is the *difference* of accelerations of the TMs. During the ascent calibration phase, we will use TFG data to guide the positioning of the TMs with the TM suspension system (TM-SS, see below) so that their differential acceleration is $< 3 \times 10^{-14}$ $g(h_m)$. Relative acceleration can be estimated to this accuracy from 3 s of TFG data.

Earth's gravity gradient $dg_z/dz$ at $h_m$ is $1.6 \times 10^{-6}$ $s^{-2}$. The Z-component of the difference of acceleration due to changes of the relative heights of the TM's is made less than $2 \times 10^{-18}$ $g(h_m)$ by using TFG measurements to guide the setting of TM relative positions to a *repeatability,* from drop to drop, of $1 \times 10^{-12}$ $m$. These measurements are with respect to the TFG plate, which has more than the stability required for this measurement. If, unexpectedly, the positions cannot be set to this accuracy, we can use the measurements for *a posteriori* correction.

Each TM comprises a pair of blocks connected by a bar displaced vertically from the center. We envision this three-part object being machined from a single block of aluminum. Each block is surmounted by a flat mirror with a gold surface that forms one end of a TFG measurement interferometer. (The rest of the TM and the surrounding walls are also gold coated, as discussed below.) The selection of test substances has been discussed in many papers [e.g., Blaser 1996]. The nominal substance pair for the first SR-POEM flight is aluminum and lead. In one TM, cylinders of aluminum are removed from the blocks and replaced by tubes of lead without changing the TM mass and having minimal effect on the moments of inertia [Reasenberg et al. 2011]. About half of the TM mass will be lead. The blocks are arranged in a square lying in a plane perpendicular to the fore-aft axis of the payload (the Z axis) and close to the CM of the free-flying payload (Fig. 2). At a distance of 0.3 m along the Z axis from the CM is the TFG plate, which is made of an ultralow expansion glass having a coefficient of thermal expansion of about $10^{-8}/K$. The TFG plate holds four concave cavity-entrance mirrors for our TFGs [Thapa et al. 2011, Phillips and Reasenberg 2005]. These mirrors have the same in-plane spacing and orientation as those on the blocks, to within approximately 100 μm.

---

[1]This approach is far simpler than a standard drag-free system because it requires neither communications between the experiment support electronics and the supporting payload buss nor low-thrust, low-noise actuators (e.g., field emission electric propulsion).

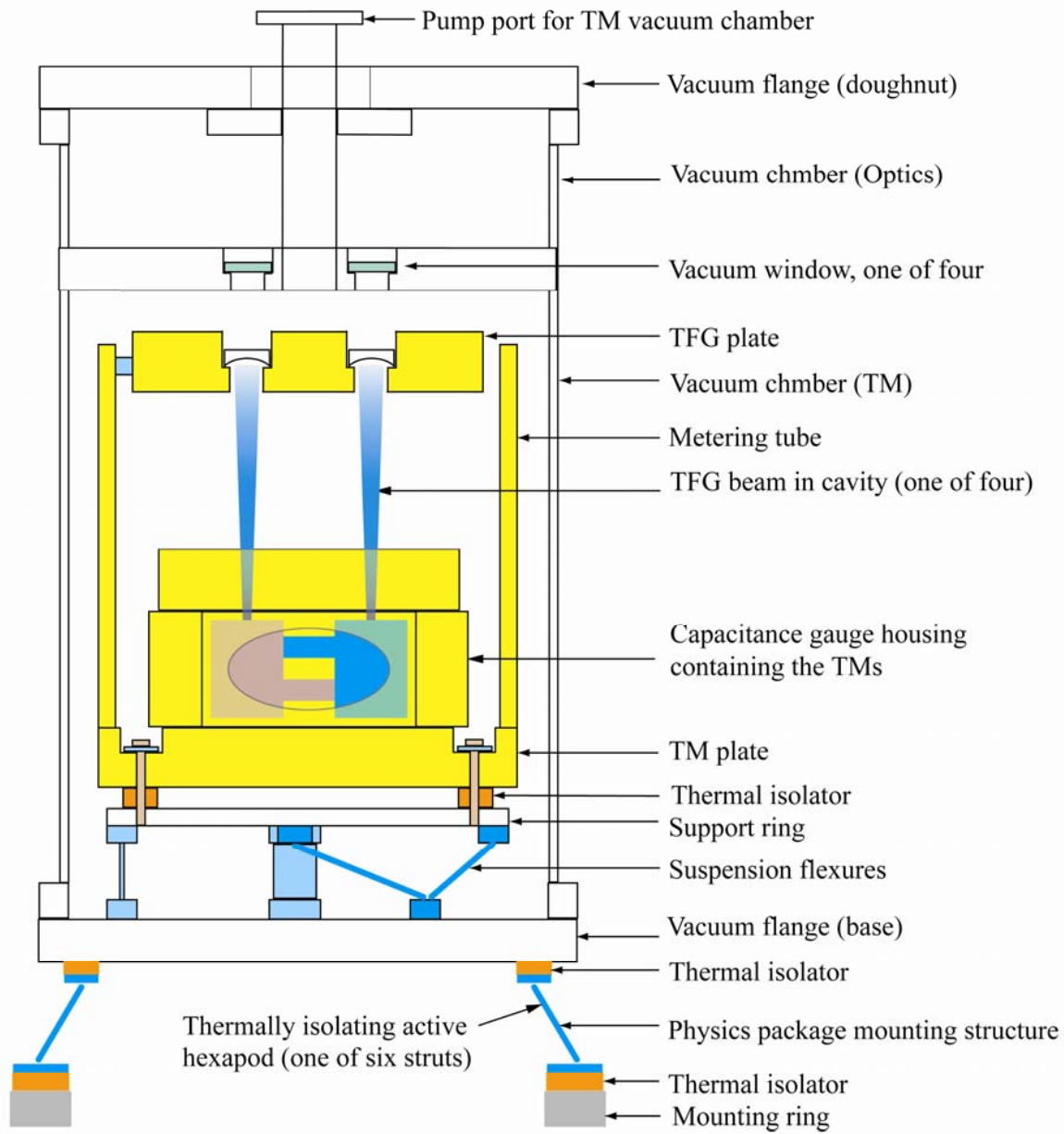

**Figure 1.** (Color online) Physics package including vacuum chamber. The precision structure has four major components (shown in yellow): 1) The TM test mass) plate holds the TM housing and provides access for the TM caging mechanism. 2) The TM housing supports the capacitance gauge electrodes; Its wall thicknesses are balanced to reduce the local gravitational gradient (gravitational spring). 3) The metering tube provides a rigid, stable connection between the TM plate and the TFG (tracking frequency laser gauge) plate. 4) The TFG plate supports the four TFG end mirrors, which are mounted over holes in the plate to allow a view of the TM. Not shown here are the vacuum port for the optics chamber, and the TFG optics. The two TMs are shown in different colors through the wall of the capacitance gauge housing. The TM chamber is at $10^{-10}$ Torr and the optics chamber is at $10^{-4}$ Torr.

The TM blocks are surrounded by the plates of a multi-component capacitance gauge that measures TM motion in all six kinematic degrees of freedom. "Drive electrodes" face the top and bottom of the TM, and the upper and lower 20 mm of the two outward-facing sides of each block. "Sense electrodes" face the central 40 mm of the outward-facing sides. The TM-housing gap is 4 mm. Pairs of drive electrodes are driven in antiphase at a frequency $f_n = f_0 + n\Delta, n = 1, ...,6$, with up to 10 V rms during setup, and 3 mV during the WEP measurement. A TM centered between electrodes driven at $f_n$ yields no signal on the sense electrodes at $f_n$. Each reading is based on data taken during an interval that is a multiple of $\Delta^{-1}$ to make the separation of signals effective. The amplitude and sign of the signal at $f_n$ indicates the departure from center. Since capacitance gauge data are taken at short intervals, their analysis can provide rates and higher time derivatives as well as positions and angles. The capacitance gauge plates can be used to apply force or torque electrostatically. This capability, when combined with the sensing, is called the TM suspension system (TM-SS), which is used both for high force applications – initial capture and slewing during payload inversion – and for precise placement. During inversions, we apply 1100 V rms, slewing 180° in 20 s.

When the sounding rocket payload reaches an altitude of about 900 km, the science-measurement phase of the flight begins. By that time, the TM have been uncaged, the TM charges have been neutralized, the Z axis has been aligned with the mean nadir direction of the pending "drop," and the TM-SS has been used to assess and then correct the position and motion of each TM in all six degrees of freedom, placing the TMs' CMs close to the payload CM, as explained above. With the preparation completed, data taken by the TFG may be used for a WEP measurement of duration $Q$. After all eight drops have been performed and before re-entry, the charge evaluation (but not neutralization) and CM position estimation performed during the ascent calibration phase are repeated.

After the flight, TFG data from the measurement cycles will be combined with auxiliary data in a weighted-least-squares fit to estimate η and its uncertainty. Several types of auxiliary data will be required. Readings from the onboard GPS will be combined with an Earth gravity model to provide the payload trajectory and the Earth's gravity vector as a function of time, likely using the SAO Planetary Ephemeris Program. To construct estimates of the physics package orientation history during the drops, we will use measurements made by the payload ACS (attitude control system) operating in "passive mode" combined with the Earth's gravity gradient matrix at the payload location, the motion history of the hexapod, and the moments of inertia and mass of the physics package supported by the hexapod and of the remainder of the payload.

The capacitance gauge data measure the orientations and the transverse positions of the TMs in the physics package frame. For each drop, the capacitance gauge drive signal is at low level during the WEP measurement period. The reduced drive level makes negligible the systematic error from drive-signal induced acceleration of the TMs. These capacitance gauge data, when combined with the TFG observations of each TM provide the information from which the real-time estimator determines the orientation and transverse position of each TM with respect to the physics package frame. This is the basis for operating the PPPS.

The data analysis is not expected to be computationally burdensome and will be done on a desk-top (PC) computer. The analysis code is straightforward, but needs to be written in a flexible manner so that diagnostics and data-quality investigations are convenient. Early runs will be used to check for and either remove or correct defective data (blunder points). The data analysis is expected to run quickly, making it easy to perform numerous numerical experiments to test for signs of systematic error.

A sounding rocket experiment is inevitably carried out in a short period of time. It must work as launched without "tuning up" by commands from a ground station. Algorithms must be robust, but not necessarily optimal. The flight software must take in stride all anomalies and yet take data on schedule. This is a challenge given the complexity of the SR-POEM event sequence, which includes pre-measurement calibration and set up, alternating measurement and payload inversion, and post-measurement recalibration. The control program (sequencer) for the mission will need to undergo extensive and realistic simulation. In addition, we plan a series of tests of the physics package onboard a "zero-g" airplane flight. This will be used to test key aspects of the mission such as uncaging, charge neutralization, operation of the PPPS, and flight software.

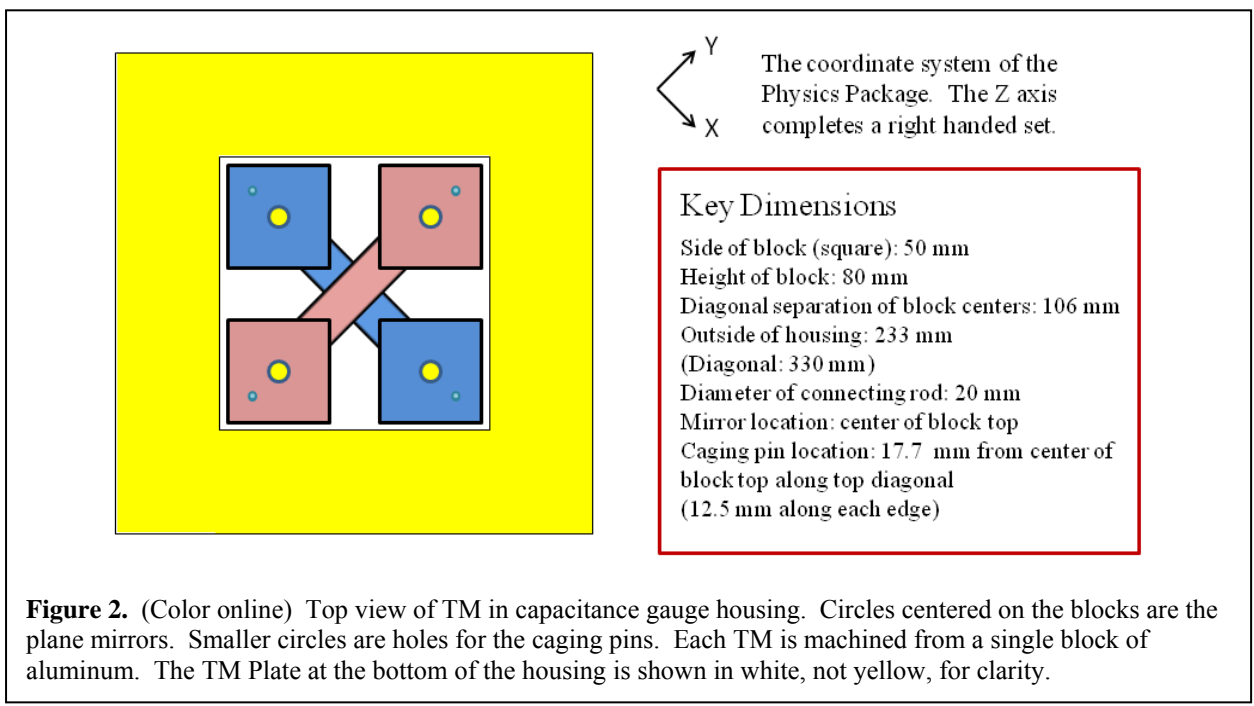

**Figure 2.** (Color online) Top view of TM in capacitance gauge housing. Circles centered on the blocks are the plane mirrors. Smaller circles are holes for the caging pins. Each TM is machined from a single block of aluminum. The TM Plate at the bottom of the housing is shown in white, not yellow, for clarity.

In the rest of Section 2, we describe the present version of SR-POEM's evolving design, with emphasis on recent changes. The original work on SR-POEM assumed that it would use one of the standard sounding rockets available at the NASA Wallops Flight Facility (WFF), particularly the most capable, the Black Brant XII. Later, we found that we needed greater capacity (mass and altitude) than the Black Brant XII affords. After some discussion with our colleagues at WFF, we tentatively settled on an Orion 50SG/Orion 38 stack, with the Orion 50SXLG (guided)/Orion 38 stack as an option for still greater altitude and mass [Eberspeaker 2011, and Brodell 2011].

*2.1. Instrument overview.*

The principal driver of the SR-POEM design is the suppression of systematic error. The CM positions of the two TMs are nominally the same, as discussed. The Z coordinate of a TM CM is obtained from the average of the two TFG measurements made between the mirrors on the two blocks and the corresponding concave mirrors mounted on the TFG plate.

The instrument precision structure is in the TM vacuum chamber, Fig. 1. It is constructed of ultralow expansion glass, and has four major components: 1) The TM plate holds the TM housing and provides access for the TM caging mechanism. 2) The TM housing supports the capacitance gauge electrodes; Its wall thicknesses are balanced to reduce the local gravitational force and force gradient (gravitational spring). 3) The metering tube provides a rigid, stable connection between the TM plate and the TFG plate. 4) The TFG plate supports the four TFG end mirrors, which are mounted over holes in the plate to allow a view of the TMs.

The associated beam steering devices and the optical detectors are housed in the optics chamber, which is isolated from the TM chamber. In the optics chamber, we can use plastic insulation and vacuum grease lubricant, since the pressure need be no lower than $10^{-4}$ Torr. Because of the nested design, external thermal perturbations are filtered with at least two time constants of more than $10^5$ s. By flying near midnight, we avoid direct solar heating,[2] which would greatly exceed other thermal perturbations and be highly directional in spacecraft coordinates. Our analysis, which also considered internal heat sources, shows no significant thermal contribution to the error in the estimate of η [Reasenberg and Phillips 2010].

The TFG optical cavities, on which are based the key distance measurements, are entirely inside the TM chamber, which is pumped down, baked and allowed to cool before launch. A small pump runs during the pre-flight and flight to maintain a pressure of $10^{-10}$ Torr, which is required to reduce Brownian force noise (see Sec. 2.7). The prime candidates are an ion pump and a sorption pump. All electrical and optical connections to the physics package are made through the end flanges of the TM and optics chambers (bottom and top in the Figure).

*2.2. Laser gauge*.

The principal measurements are made by the four TFGs observing the two TMs. In operation, each block is continuously observed. Because of its high precision, 0.1 pm in 1 s, the TFG allows a high accuracy test of the WEP during the brief period afforded by a sounding rocket. The TFG's speed of measurement, coupled with the payload inversions, shifts upward the frequency band of interest for the measurements to 0.003 Hz and above. This shift reduces susceptibility to systematic error.

The SR-POEM measurement will employ four Fabry-Perot cavities, each with a curved mirror at the TFG plate and a plane gold-coated mirror on one of the TM blocks. The optical beam will be injected through the curved mirror and kept aligned by an automated system using Anderson's method [1984], which was demonstrated by Sampas and Anderson [1990] and which has recently been shown to work in a reflection cavity [Reasenberg 2012]. In order to use Anderson's method in reflection, we give the curved mirrors a higher reflectivity than gold (97.5±1%) at our laser wavelength of 1550 nm.

Our current laser gauge implementation, and the one for SR-POEM (Fig. 3), is the Semiconductor Laser TFG. In it, the Variable Frequency Source comprises a tunable distributed feedback (DFB) semiconductor laser operating at 1550 nm and a phase modulator. The Variable Frequency Source's optical frequency is locked to the Fabry-Perot cavity. To measure position changes of a TM, we measure changes in the optical frequencies of its two TFGs. For all four TFGs, there is one reference laser, which is locked to a very stable cavity or to an atomic line. A portion of the light from each TFG's tunable laser is heterodyned against a portion of the reference laser light to provide a radio frequency output. Measuring this radio frequency provides the needed measure of optical frequency change, and thus of distance change. The WEP observable is the difference of the averaged output from one TM and the average output from the other TM. The SR-POEM configuration has four TFGs whose paths are matched to within 100 μm. Because the measured paths are nearly equal, the stability of the reference laser

---

[2] Even at midnight, the Sun can be seen at the summer solstice (12.5 degrees above the horizon, including 1.1 deg of refraction) from 2000 km above the Wallops Flight Facility. At the winter solstice, an altitude over 16,000 km is needed to see the Sun.

frequency is relatively noncritical for the WEP data. However, reference laser drift also causes a drift in the feedback signal for the PPPS. A tolerance of 3 nm on drift of TM position implies a requirement

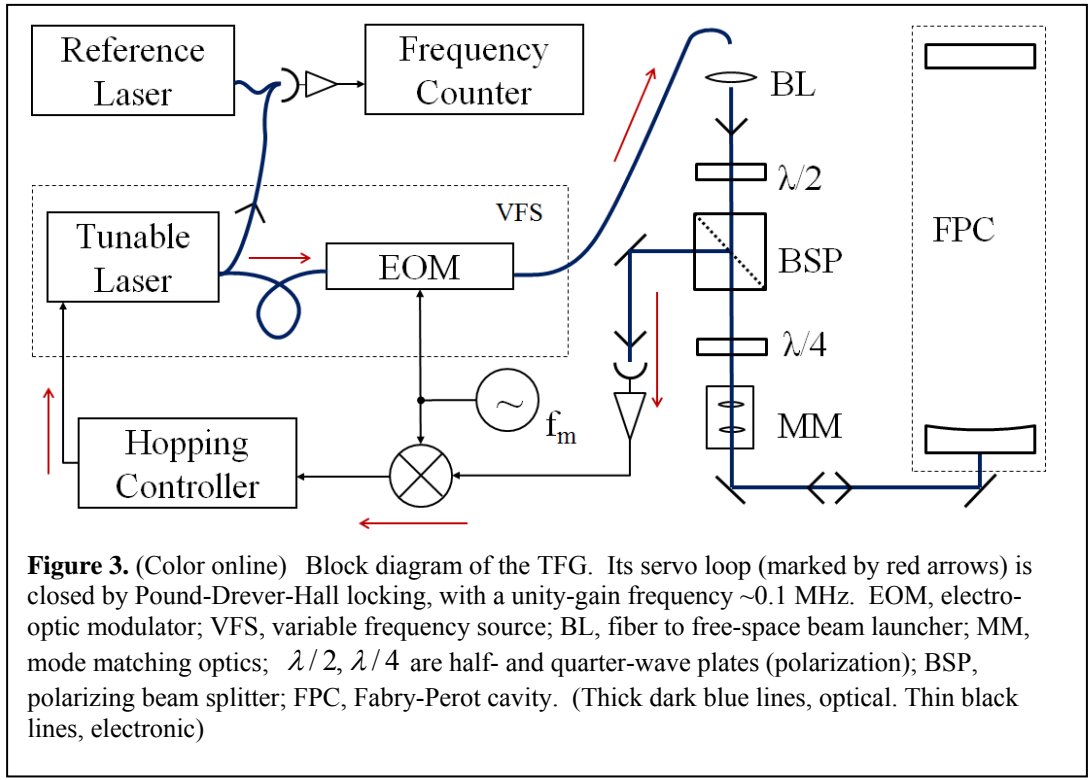

**Figure 3.** (Color online) Block diagram of the TFG. Its servo loop (marked by red arrows) is closed by Pound-Drever-Hall locking, with a unity-gain frequency ~0.1 MHz. EOM, electro-optic modulator; VFS, variable frequency source; BL, fiber to free-space beam launcher; MM, mode matching optics; $\lambda/2$, $\lambda/4$ are half- and quarter-wave plates (polarization); BSP, polarizing beam splitter; FPC, Fabry-Perot cavity. (Thick dark blue lines, optical. Thin black lines, electronic)

for reference laser stability of 2 MHz. This can be exceeded tenfold by using for the reference a laser locked to a Fabry-Perot resonator in optical fiber made with temperature-compensated fiber, whose temperature is kept stable to 1 mK.

The SR-POEM requirement for the TFG is an uncertainty in distance of 0.1 $pm\ Hz^{-1/2}$ for periods up to 300 s. The TFG sensitivity is computed as the Allan deviation

$$z(\tau) = \left[ \frac{1}{2(N-1)} \sum_{i=1}^{N-1} (y_{i+1} - y_i)^2 \right]^{1/2} \quad (4)$$

where

$$y_i = \frac{1}{\tau} \int_{(i-1)\tau}^{i\tau} y(t) dt \quad (5)$$

and $z(\tau)$ is a measure of the repeatability of the measurement on a time scale $\tau$. The achieved performance and the requirement are shown in Fig. 4. A principal error source, spurious amplitude modulation introduced by partially-reflecting components in the optical path from the Variable Frequency Source to the measurement interferometer (FPC in Fig. 3), has been controlled to date with optical isolators in the fiber portion of the path. We anticipate that at least part of the required improvement will be achieved by adding a monitor for spurious amplitude modulation, allowing active suppression via feedback to an amplitude modulator.

*2.3. TM rotation and Beam alignment.*

For a body in free fall near Earth, with one principal axis (*a*) horizontal and another making an angle $\mu_a$ with the vertical, the Earth's gravity gradient causes an angular acceleration around the *a* axis given by

$$\ddot{\mu}_a = \frac{3g(r)}{r}\frac{I_c - I_b}{I_a}\sin(\mu_a)\cos(\mu_a) \tag{6}$$

where ($I_a$, $I_b$, $I_c$) are the principle moments of inertia around axes a, b, and c, respectively. If the body is a thin rod ($I_c = 0$) and µ is always small (i.e., the rod is near vertical and librating), then it will have a pendulum period $P_{rod} = P_{orbit}/\sqrt{3}$, where $P_{orbit}$ is the period of a circular orbit at the rod's altitude. In our case, $P_{rod} \approx 4\ 10^3$ s.

Imagine a thin rod in place of the SR-POEM payload and consider its motion for the duration of a drop. We take µ(t) to be the angle between the thin rod at time *t* and the vertical at a time $t_{1/2}$ that is half way through the drop. If the rod is nearly inertial, then in the orbital plane it makes an angle $\varphi$ with the local vertical: $\varphi = \mu(t) - \dot{v}(t - t_z)$. Ideally, $t_z = t_{1/2}$ and $\dot{v}$ is the payload's orbital angular rate (treated here as if constant), where v is the payload's true anomaly. Here we will take $t_z - t_{1/2} = 0$ and $\dot{v} =$ the Earth rotation rate of about 15 arc sec / s. Thus, by the design of the mission, the gravity gradient torque goes to zero and then reverses at $t_{1/2}$. With proper initial conditions, the maximum angular excursion $\tilde{\mu}$ of the rod during a free-fall period can be made small. For Q = 120 s, making $\mu = \tilde{\mu}$ with the correct initial velocity at the start (and end) of the drop, such that µ is zero in the middle of the drop, yields $\tilde{\mu} \approx 2 \times 10^{-6}$. For the more nearly spherical payload, $\tilde{\mu}$ would be smaller. The two TM and the rest of the payload will each have a unique value of the ratio of moments of inertia used in Eq. 6. Thus, even if the TMs are not initially rotating with respect to the mean orientation of the payload, all three will rotate apart. These motions alone would cause the cavity to become misaligned by the order of $2 \times 10^{-6}$ radian in the absence of an alignment servo.

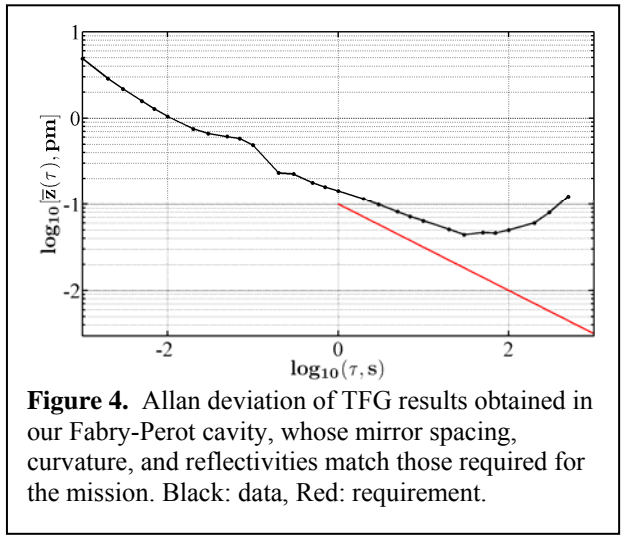

**Figure 4.** Allan deviation of TFG results obtained in our Fabry-Perot cavity, whose mirror spacing, curvature, and reflectivities match those required for the mission. Black: data, Red: requirement.

We address this problem of optical cavity misalignment at two levels. First, the alignment servo for each TFG maintains accurate alignment to the $TEM_{00}$ mode, but results in "beam walk," s, on its curved mirror with peak value well under 1 µm, which should be compared to the spot size, $\omega = 700$ µm. Because of optical imperfection, beam walk can cause an error in the measured distance. For example, if the mirror radius has an error u (i.e., $R = R_0 + u$, where $R_0$ is the intended radius that puts the center of curvature at the height of the TM center of mass), then the distance measured by the TFG has an error $\epsilon = u(1 - \cos\theta)$, where $\theta = s/R$ is the beamwalk angle. For R = 0.25 m, u = 1 mm and s = 1 µm, $\epsilon = 8 \times 10^{-15} m$, which should be compared to the distance sensitivity of the TFG during a drop: $0.1\ pm/\sqrt{120} = 9 \times 10^{-15} m$. $\epsilon$ will be smaller, on average, and will not have a parabolic signature. The smallest detectable parabolic signature is $2.3 \times 10^{-13} m$ ($6.5 \times 10^{-13} m$), which corresponds to the intended mission (single drop) sensitivity. Finally, if $\theta$ is biased, $\theta = \theta_0 + s/R$, then there is an additional contribution to the distance error, but it is antisymmetric in time, and makes no contribution to the sought after parabolic signature.

**Table 1**. PPPS degrees of freedom (DOF) and error signals.

| DOF | Error signal |
|---|---|
| X-,Y-rotation | Difference of TFG readings for corresponding TM |
| Z-rotation | Capacitance gauge |
| X-, Y-position | Capacitance gauge |
| Z-position | Average of all four TFG readings |

Second, TFG and capacitance gauge readings are used to drive the PPPS, as detailed in Table 1. The dominant non-gravitational force is radiation pressure on the payload from a warm Earth, which causes acceleration of the payload, including the physics package, with respect to the (approximately inertial) average frame of the two TMs. The PPPS moves the physics package with respect to the rest of the payload, nulling the motion of the physics package with respect to the TMs. For Z-position, the PPPS error signal from the TFGs is far better than required. The noise is 0.1 pm $Hz^{-1/2}$; and the requirement is 200 nm over a drop.

Before each drop, the payload is oriented and the TM-SS used to set the initial conditions, which ideally would be: the CMs of the two TMs coincident and comoving with the payload CM; the physics package inertially pointed; and the TM oriented toward the TFG and rotating slowly with respect to the rest of the physics package as required to achieve the intended small $\tilde{\mu}$. The beam alignment system provides ancillary information about the relative orientation of the experiment and TM by monitoring the position on the input mirror where the TFG beam is injected.

Normally, one considers four degrees of alignment freedom per cavity, two positions and two angles. However, the only motion of cavity elements that needs real-time correction is a rotation of each TM about its CM. This can be corrected by adjusting the angle of the beam entering the cavity as seen from the flat mirror, rotating it about a point P, at the intersection of the optical axis with the X-Y plane containing the TM's CM. In the optical train that excites the cavity, there is a "tip-tilt mirror" whose surface is imaged onto the X-Y plane at P. This mirror is driven through a suitable filter-amplifier by the errors sensed by Anderson's method.

*2.4. Coriolis acceleration and spacecraft ACS (attitude control system)*

Nominally, during the WEP measurement, the payload is inertially pointed and the TMs have zero transverse velocity in the coordinate system of the physics package. In practice, both the rotation and transverse motion will be non-zero. Thus, we expect some Coriolis acceleration for each of the TM. Since this acceleration depends on quantities that are random and unlikely correlated between drops, we can expect no cancelation of the Coriolis acceleration from the payload inversions. We can, however, keep the rotation and transverse motion small, measure these small motions precisely, and use these data to model (correct for) the pseudo-acceleration.

In preparation for a drop, the following six steps are taken: 1) The payload is brought to a near vertical orientation such that over the duration of the drop the average orientation is vertical and the average three-axis rotation rate is zero. 2) The ACS is transitioned from active to passive; the sensors, a fiber-optic gyro and a star tracker, keep taking data but there is no actuation. 3) The TMs are brought to the starting position and velocity in the physics-package frame by the TM-SS. 4) The TM-SS is disabled (put in passive mode). 5) The capacitance gauges provide a precise determination of the TM locations (in X and Y) by running briefly at high excitation signal. 6) The exciting signal on the capacitance gauges is made small, putting the TM-SS in "quiet mode." Following the WEP observation of duration $Q$, the capacitance gauge exciting signal is again increased briefly to measure the precise locations in the X-Y plane of both TMs. This ends the drop. During the payload inversion that follows the drop, the TM-SS again controls each TM's six degrees of freedom but there is no recaging.

The startracker data gathered during the drop provide the basis for determining the payload rotation history. Since the ACS actuation is disabled, there is a simple physical model that can be fit to the data. Similarly, the mean velocity of the TMs in the frame of the physics package is easily found from the precise position data taken just before and just after the collection of the WEP data. Here we consider the uncertainty in the estimate of the differential Coriolis acceleration. Let $\varsigma$ be the Z component of the differential TM Coriolis acceleration. Then

$$\sigma(\varsigma) = 4\,[\sigma^2(\delta\hat{\omega})\sigma^2(\delta v) + \sigma^2(\delta\omega)\,\sigma^2(\delta\hat{v}) + \sigma^2(\delta\hat{\omega})\,\sigma^2(\delta\hat{v})]^{1/2} \qquad (7)$$

where $\delta v$ is the error in zeroing one component of transverse velocity of one TM at the start of the drop, $\delta\omega$ is the error in zeroing one component of the rotation of the payload around an axis in the X-Y plane, and $\delta\hat{v}$ and $\delta\hat{\omega}$ are the estimates of those quantities based on data accumulated by the end of the drop. Anticipating the results below, we note that of the three terms in Eq. 7, the first two are of nearly the same order and the last is much smaller. We next estimate the four uncertainties in Eq. 7.

$\delta v$ is determined as the TMs are being setup and is limited by the precision of the capacitance gauge. We make the drive signal large during this period, $E_{CG}$ = 10V rms. Then,

$$\sigma(v) = 1.4\times10^{-11} m/s \quad \frac{\sigma_{CG}}{10^{-9} Volt/\sqrt{Hz}}\frac{d}{4\,mm}\frac{0.1}{D}\frac{10\,Volt}{E_{CG}}\left(\frac{1\,s}{T_{CG}}\right)^{3/2} \qquad (8)$$

where D is the (unavoidable) capacitive voltage divider before the first amplifier and d is the TM-housing distance. For a capacitance-gauge electronic sensitivity of $\sigma_{CG} = 10^{-9} Volt/\sqrt{Hz}$ and an observing time of $T_{CG}$ =0.25 s, the velocity uncertainty is about $\sigma(v) = 290$ pm/s. Similarly, the capacitance gauge measures position with precision

$$\sigma(x) = 4\times10^{-12} m \quad \frac{\sigma_{CG}}{10^{-9} Volt/\sqrt{Hz}}\frac{d}{4\,mm}\frac{0.1}{G}\frac{10\,Volt}{E_{CG}}\left(\frac{1\,s}{T_{CG}}\right)^{1/2} \qquad (9)$$

which yields a position uncertainty of 8 pm in $T_{CG}$ =0.25 s. Since we have such a measurement both before and after the 120 s WEP measurement, the uncertainty at the end of the drop of the mean velocity during the drop is $\sigma(\delta\hat{v}) = 0.094\,pm/s$.

We may apply the same kind of analysis to the payload rotation. The precision pointing system for a sounding-rocket payload uses an ST5000 startracker from the Space Astronomy Laboratory of the University of Wisconsin-Madison. It has a measurement uncertainty of 0.8 arc-sec in 0.1 s and successive determinations are statistically independent [Percival et al. 2008]. Thus, $\sigma_0$ = 0.25 arc-sec (i.e., in 1 s). For now, we neglect the more precise but likely biased short-term behavior of the gyro, which is expected to be part of the ACS. Then the ACS can determine angle rate with precision

$$\sigma(\omega) = \frac{0.866\,arc\,\mathrm{sec}}{T_{ST}}\sqrt{\frac{1\,s}{T_{ST}}} \qquad (10)$$

If the pre-drop rotation rate of the payload is determined based on $T_{ST}$ = 1 s of data, then $\sigma(\delta\omega) = 0.9\,arc\,\mathrm{sec}/s$. After 120 s of WEP measurement, $\sigma(\delta\hat{\omega}) = 6.6\times10^{-4}\,arc\,\mathrm{sec}/s$.

By using the above error estimates, we can find the magnitude of the Coriolis acceleration, $5\times10^{-15} ms^{-2}$, which should be compared to $3\times10^{-16}\,m/s^2$, the per-drop error that averages to the stated mission accuracy after eight drops. Thus the Coriolis acceleration must be computed and included in the mission analysis. We can also find the uncertainty in the estimate of the Coriolis acceleration based on data available by the end of a drop: $\sigma(\varsigma) = 4.2\times10^{-18}\,m/s^2$, for $Q = 120s$. This is two orders of magnitude smaller than the intended per-drop accuracy. Since the Coriolis error is expected to be random, it will average over eight drops to $1.5 \times 10^{-18} \, m/s^2 \approx 2.3 \times 10^{-19} \, g(h)$.

*2.5. WEP measurement sensitivity.*

If a single laser gauge has distance measurement uncertainty $\sigma_0$ for an averaging time $\tau_0$, then an analysis of many of its measurements with total time $T$ (assuming white noise and $\tau_0 \ll T$) results in an estimate of acceleration with uncertainty

$$\sigma_{acc} = \frac{\sigma_0(\tau_0)}{Q^2} \sqrt{\frac{\tau_0}{T}} K \tag{11}$$

where $Q$ is the single-drop free-fall time, and $K = 12\sqrt{5} \approx 27$, assuming that position and velocity are also estimated. The corresponding WEP sensitivity for the pair of TMs and four TFGs is $\sigma(\eta) = \sigma_{acc}/(R\,g(h))$, where $R$ is the fraction of TM mass that is test substance, $g(h)$ is the gravitational acceleration, and $h$ is the altitude of the instrument. In a reasonable scenario, $\sigma_o = 0.1\,\text{pm}$ (for $\tau_0 = 1s$), $Q = 120$ s, $R = 0.5$, and there are eight measurements (separate drops) so that $T = 960$ s. This yields $\sigma(\eta) = 1.7 \times 10^{-18}$ based solely on TFG noise.

*2.6. Relationship between $\sigma(\eta)$ and maximum altitude*

The TFG contribution to the uncertainty in the estimate of η is proportional to $Q^{-2}T^{-1/2}$. Both $Q$ and $T$ are limited by the free-fall time afforded by the sounding rocket. We can improve $\sigma(\eta)$ by increasing $Q^2\sqrt{T}$, as long as systematic error is correspondingly reduced. If we decrease the number of drops from eight to six, the increase in $Q$ and slight increase in T significantly increases the precision of the acceleration measurement. However, at present we regard the larger number of drops as required, for the increased robustness of the experiment that it conveys. We therefore consider the effect of increasing the apogee of the payload orbit. The increased apogee gives more free-fall time, but with a decreased average value of gravitational acceleration.

If the extra free-fall time from increased altitude were used to allow extra drops of the nominal duration, then doubling the altitude from the old nominal of 1200 km would lower $\sigma(\eta)$ by only 20%, an uninteresting achievement. Here we assume the extra time will be used to increase $Q$, that the calibration time, starting at an altitude of 200 km, will be held at 150 s, and that the time required to invert the payload and prepare for the drop is held fixed at 30 s.[3] We take into consideration the declining value of g(h) as altitude, h, increases and ignore systematic effects. Then the contribution of the laser gauge measurement uncertainty to the total mission error is shown in Fig. 5. With eight drops, a maximum altitude of 2000 km is sufficient for a mission that yields $\sigma(\eta) = 10^{-17}$. A higher flight might be necessary if we find it will take more than 150 s to do the setup and ascent calibration.

---

[3] The inversion requires 20 s according to the attitude-control engineers at WFF. The remaining 10 s is used to bring the TM to their nominal starting positions and zero velocity in the payload frame. Note that the TM-SS has a unit-gain frequency of at least 10 Hz.

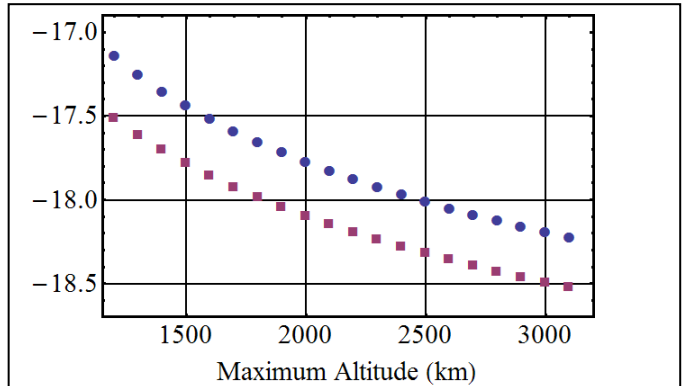

**Figure 5.** (Color online) Log of the contribution to the uncertainty in the measurement of η from the laser gauge measurement of TM acceleration. Circles (blue), 8 drops per flight; squares (red), 6 drops per flight. Setup and calibration kept at 150 s starting at an altitude of 200 km.

*2.7 Electrostatic force*

As the TM is freed from its launch restraint, it will acquire a potential of a few × 0.1 V. This is easily measured by briefly applying a constant potential to the capacitance gauge electrodes and using the TFGs to determine the acceleration. We plan to neutralize the TMs iteratively by illuminating each TM and one of its surrounding (TM-SS) electrodes with an ultraviolet light-emitting diode, and biasing the charge flow by putting a potential on the illuminated electrode [Sun 2009]. Some of the remaining electric field, that due to the difference of average surface potential of the TM top and bottom faces, will be made zero by applying a matching potential to the corresponding TM-SS electrodes.

The more difficult problem comes from the spatial variation of potential across a single face, often called the patch effect [see Camp et al. 1991]. A standard solution is to use a gold surface and to ensure that it is clean. The gold is often deposited over another thin film of material (e.g., germanium) that will make the gold deposit more uniformly and adhere well. An extensive Kelvin-probe study by Robertson et al. [2006] examined numerous surface-coating combinations, finding spatial fluctuations of potential of 1 to 2 mV rms with respect to the mean on the "best" surfaces, which included surfaces with gold as a top layer. These results were at the discretization level of their Kelvin probe. Robertson et al. found additionally that the quality of the vacuum made a large difference to the results. Camp et al. [1991] also found that their best surfaces, including gold and graphite, had a variation of 1 mV rms, which was at the limit of resolution of their instrument. We are developing a Kelvin probe with higher spatial resolution and much lower surface potential noise (including discretization).

The above spatial fluctuations of potential would be acceptable as long as the potential is sufficiently constant. We set a requirement on constancy during inversions. Let $V_d(x, y, t)$ be the potential on the surface of the TM or its housing, spatially averaged to remove features of spatial frequency $> 1/d$, where $d = 4\ mm$ is the TM-housing distance. Features of higher frequency contribute unimportantly to the force, which results only from field lines that cross the gap. We assume that the change in $V_d$, i.e., $V_d(x, y, t) - V_d(x, y, 0)$, is not correlated with $V_d(x, y, 0)$. Let $f(t) = \overline{V_d(x, y, t)^2}^{1/2}$, where the mean is taken over the TM or housing surface. We require that the root power spectral density of $f(t)$ at the 3.3 $mHz$ inversion frequency be less than $1 \times 10^{-3}\ V/Hz^{1/2}$. Then the change in differential acceleration is less than $2 \times 10^{-18}$ g(h), which is acceptable. Robertson *et al,* [2006] give the root power spectral density for a single one of their samples, for 2400 s of data taken at a single position on one sample. The root PSD is quite flat over the entire frequency range plotted, which includes 3.3 $mHz$. It has a value of $1 \times 10^{-3}\ V/Hz^{1/2}$.

Recent work by Pollack et al. [2008] looks directly at the force between a pair of surfaces using a torsion balance, but the force noise of the torsion balance, which appears to come from the torsion fiber and from residual gas in the chamber, is an order of magnitude greater than the level we require. Pollack et al. also looked at the variation of average potential, and find that for frequencies above 0.1 mHz there is a white potential-fluctuation spectrum with density $30\mu V/\sqrt{Hz}$, which is 1.5 orders smaller than the requirement. However, there is very little published information on time variation of surface potential. We must have more information on time variation on the surfaces we intend to use, on spatially-varying time variation, and on the correlation of spatially-resolved temporal variation with the initial spatial variation. The Kelvin probe measurements that we are planning, which will have improved resolution for potential and for spatial and temporal variations, will provide this information.

*2.8. Magnetic force and shielding*

A TM with a dipole moment *m* in a magnetic field *B* will experience a force

$$\vec{F} = \vec{\nabla}(\vec{m} \cdot \vec{B}) \tag{12}$$

Reported magnetic moments for samples of aluminum and similar metals, when scaled to the size of the SR-POEM TM, yield m = 5.0 $10^{-8}$ A m$^2$. [Patla et al. 2012, in preparation] The induced moment is expected to be 100 fold smaller and will not be discussed further. Magnetic shielding in a tight space is inefficient. Further, some components of the magnetic force do not cancel in an inversion. Guided by the shielding equations given by Sumner et al. [1987], we evaluated several configurations using finite element analysis.

All shielding designs investigated to date have been to fit in the small space afforded by the Black Brant XII launch vehicle. The best among these have either three closely spaced concentric shields or two closely spaced shields and a field-canceling solenoid, which saves weight. Of the designs that had the required shielding factor, the lowest mass design had two shields and a solenoid and a mass of 15 kg, exclusive of the electronics to drive the coil. [Patla et al. 2012, in preparation]

With the larger fairing of the Orion 38 second stage, we can design a more efficient shielding system by having more space between the outer and inner shields. We assume a ten-fold improvement in shielding, yielding a gradient of $\leq 1 \times 10^{-10}\ T/m$. The contribution to differential TM acceleration is $\leq 2 \times 10^{-18}\ g(h)$. Further, we will make the outer shield a closed cylinder with a diameter to height ratio that behaves like a spherical shield and nulls the torque on the shield from the ambient field.[4] This is important since we plan to disable the ACS (attitude control system) actuation during the drops and a shield that is far from spherical could produce an angular acceleration that is a significant fraction of 10$^{-5}$ s$^{-2}$.

*2.9. Brownian motion noise*

Recently, the increase of Brownian force noise due to residual gas when surfaces are close to a test mass has been discovered [Cavalleri 2009] and quantified [Schlamminger 2010, Dolesi 2011]. To make an accurate estimate of this effect requires a Monte Carlo calculation for the specific geometry. Working by analogy with the Monte Carlo calculations that have been made for other geometries and validated by experiments, we have made what we believe is a conservative estimate that, with the gas pressure around the TMs reduced to 10$^{-10}$ Torr, this acceleration noise after is only 0.16 of the intended SR-POEM uncertainty. The actual SR-POEM geometry, for which no Monte Carlo calculation has yet been done, may reduce this noise. Consider the four vertical faces of a TM block. Only two of these four are close to the housing (Fig. 2). Gas is relatively free to travel vertically past the other two, which reduces the Brownian noise. After we have quantified this noise for our geometry, we may be able to relax the vacuum requirement.

---

[4] A long thin shield will tend to point along the field and a short wide shield will tend to point (the symmetry axis) perpendicular to the field. Finding the ratio of diameter to height requires finite element analysis.

*2.10. Gas pressure*

Consider a 10% imbalance of gas pressure between the top and bottom surfaces of a TM in the $10^{-10}\ Torr$ pressure of the TM chamber. This causes an acceleration of $1 \times 10^{-12}\ g(h)$. Changes in this value must be suppressed by a factor $5 \times 10^5$ to reduce the error to $2 \times 10^{-18}\ g(h)$. Gas evaporation rates vary with time due to the approach to equilibrium and to variations in temperature. The TM chamber will be baked at 100-150°C for at least several days prior to launch. The period for cooling and coming to thermal equilibrium is likely to be weeks. During the cooling of the interior, the temperature of the outside of the chamber is held constant with flowing gas whose temperature is regulated. During the late stage cooling, the regulation is to ~1 mK. The launch is near midnight in a season in which the Sun is not seen directly at any point on the trajectory.

The thermal isolation of the experiment is aided by two filters with time constants of at least $10^5$ s. The inner filter is provided by gold-coating the outside of the vacuum chamber. Creating a long thermal time constant inside the chamber is undesirable because it will slow the cooling after the chamber bake out. That cooling will take place in air, making the emissivity of the outside of the chamber relatively unimportant. In the previous design, the second thermal filter of external heat was the 17 inch tube in which the instrument was housed. In the current design, the location of the outer thermal shield has not yet been selected.

The 1200 s of the science portion of the flight is quite short by comparison. The temperature of the metering structure is expected to be constant to within a few μK over the course of the experiment. The component of temperature variation in synchronism with inversions will be smaller. This will achieve the required suppression of variation of gas pressure.

*2.11 Uncaging*

The TMs need to be caged during launch because of the high levels of vibration and upward acceleration from the engines. Since the experiment requires that the TMs be in a UHV environment, the caging pins and the TMs will tend to stick together during uncaging. LISA has approached this problem with a three-stage uncaging system [Jennrich 2009]. We are investigating a single-stage uncaging system made possible by reducing the adhesion of the pins and the TMs, and by employing a higher electric field both to pull the TM from the pins and to stop the TMs after they are pulled free. One possibility is to make the pins of SiC. Au is known to wet SiC poorly, i.e., tends not to stick. A second possibility is to use aluminum coated with gold, and on the gold to grow a self-assembled monolayer of an alkanethiol such as of *n*-docosanethiol, $CH_3(CH_2)_{21}SH$. Even after being pressed together hard enough to cause plastic deformation of the gold, such a surface has been shown not to stick to gold [Thomas et al. 1993]. Contact between the caging pins and the TMs will be at the bottom of small holes in the TMs so as to minimize the electrostatic force due to surface potential changes caused by the pins.

## 3. Conclusion

We are developing a Galilean test of the WEP, to be conducted during the free-fall portion of a sounding rocket flight. The test of a single pair of substances is aimed at a measurement uncertainty of σ(η) < 2 × $10^{-17}$ after averaging the results of eight separate drops during one flight. We have investigated sources of systematic error and find that all can be held to well below the intended experiment accuracy.

At the core of the mission design is the mitigation of systematic error, which is supported by the three stages of measurement differencing. First, the TM positions are each measured by laser gauges with respect to the commoving physics package. These measurements contain the acceleration of the physics package (principally from warm-Earth radiation pressure) and initial-velocity errors. The changes in the lengths of the measurement paths over the 120 s drop period would be at the sub-micron level, but are reduced to the sub-nanometer level by the PPPS (physics package position servo). Second, those measurements are differenced, removing the residual physics-package acceleration. Third, the payload is inverted and the differential accelerations are differenced. This last step adds the WEP violating signals and subtracts the accelerations of the TM due to fixed electrostatic, gas pressure and radiation pressure forces, and to local gravity and the symmetric part of the Earth's gravity gradient. (The next, asymmetric term is both easily calculated and below the threshold of interest.)

Ideally, at the start of a drop, the CM of each TM should be at the CM of the payload. There will be an error in this setting due to the uncertainty in the onboard calibration (determination of the position of a TM such that its CM is at the CM of the payload), which will be much better than the corresponding ground-based measurement. However, the initial conditions for the drop will be highly reproducible because the key position (along the Z axis) is sensed by the laser gauge. Thus, the inversions will cancel the error in $\hat{\eta}$ due to an error in the location of the TM at the payload CM.

A successful SR-POEM Mission would provide a 10,000 fold increase in the accuracy of our knowledge of the WEP violation parameter, η. Either the discovery of a violation of the WEP or its deeper confirmation would inform the development of gravitational physics.

## Acknowledgements

This work was supported in part by the NASA Science Mission Directorate through grants NNC04GB30G and NNX07AI11G (past) and NNX08AO04G (present). Additional support came from the Smithsonian. We thank the staff at the Wallops Flight Facility for their generous contribution to our understanding of the capabilities and constraints of a sounding rocket flight.

## References

Anderson, D.Z., "Alignment of resonant optical cavities," Applied Optics, **23**(17), 2944-2949 1984.

Berg, E.C., Bantel, M.K., Cross, W.D., Inoue, T., Newman, R.D., Steffen, J.H., Moore, M.W., Boynton, P.E., "Laboratory Tests of Gravitational Physics Using a Cryogenic Torsion Pendulum," Proc. 10$^{th}$ Marcel Grossman Meeting, Rio de Janeiro, Brazil, 20-26 July 2003, Eds.: Mário Novello; Santiago Perez Bergliaffa; Remo Ruffini. Singapore: World Scientific Publishing, p. 994 (2005).

Blaser, J.P., "Test mass material selection for STEP," *Class. Quant. Grav.*, **13**, A87-A90, 1996.

Brodell, C., Wallops Flight Facility, private communication, 2011.

Camp, J.B., T.W. Darling, et al., "Macroscopic variations of surface potentials of conductors," *J. Appl. Phys.*, **69,** 7126-7129, 1991.

A. Cavalleri, "Increased Brownian Force Noise from Molecular Impacts in a Constrained Volume," *Phys. Rev. Lett.* **103**, 140601 (2009).

Chhun, R., D. Hudson, P. Flinoise, M. Rodrigues, P. Touboul, B. Foulon, "Equivalence principle test with microscope: Laboratory and engineering models preliminary results for evaluation of performance," *Acta Astronautica*, **60**, 873-879 (2007).

R. Dolesi, et al., "Brownian force noise from molecular collisions and the sensitivity of advanced gravitational wave observatories," *Phys. Rev.* **D84**, 063007 (2011).

Everitt, C.W.F. et al., "Gravity Probe B: Final Results of a Space Experiment to Test General Relativity," *Phys. Rev. Lett.*, 106, 221101 (2011). See also http://einstein.stanford.edu/.

Eberspeaker, P., Wallops Flight Facility, private communication, 2011.

Fischbach, E. and C.L. Talmadge, *The Search for Non-Newtonian Gravity*, Springer-Verlag, New York, 1999.

Hammond, G.D., C.C. Speake, C. Trenkel, and A.P. Paton, "New Constraints on Short-Range Forces Coupling Mass to Intrinsic Spin," *Phys. Rev. Lett.*, **98**, 081101 (2007).

Jennrich, O., "LISA technology and instrumentation," Class. Quant. Grav., **26**, 153001, (2009).

Johns Hopkins University Applied Physics Lab, Space Dept., and the Guidance & Control Lab of Stanford University, "A Satellite Freed of all but Gravitational Forces: 'TRIAD I'." *J. Spacecraft and Rockets*, Vol . **11**, No, 9, pp. 637-644 (1974).

Mester, J.C. and J.M. Lockhart "Remanent Magnetization of Instrument Materials for Low Magnetic Field Applications." *Czechoslovak Journal of Physics, Supplement S5, Proceedings of the 21st International Conference on Low Temperature Physics, Prague, August 8-14, 1996* **46**: 2751-2752 (1996).

Newman, R., "Prospects for terrestrial equivalence principle tests with a cryogenic torsion pendulum," *Class. Quant. Grav*., **18**, 2407-2415 (2001).

Nobili, Anna M.; Comandi, Gian Luca; Doravari, Suresh; Bramanti, Donato; Kumar, Rajeev; Maccarrone, Francesco; Polacco, Erseo; Turyshev, Slava G.; Shao, Michael; Lipa, John; Dittus, Hansjoerg; Laemmerzhal, Claus; Peters, Achim; Mueller, Jurgen; Unnikrishnan, C. S.; Roxburgh, Ian W.; Brillet, Alain; Marchal, Christian; Luo, Jun; van der Ha, Jozef; Milyukov, Vadim; Iafolla, Valerio; Lucchesi, David; Tortora, Paolo; de Bernardis, Paolo; Palmonari, Federico; Focardi, Sergio; Zanello, Dino; Monaco, Salvatore; Mengali, Giovanni; Anselmo, Luciano; Iorio, Lorenzo; Knezevic, Zoran, "'Galileo Galilei' (GG) a small satellite to test the equivalence principle of Galileo, Newton and Einstein," *Experimental Astronomy*, **23**, 689-710 (2009).

Overduin, J., F. Everitt, J. Mester, P. Worden, "The Science Case for STEP," *Advances in Space Research*, **43**(10), 1532-1537 (2009).

Patla B., E. Lorenzini, J.D. Phillips, and R.D. Reasenberg, "SR-POEM requirements for spurious acceleration reduction," BAPS.2010.APR.S10.6.

Percival, J.W., K.H. Nordsieck and K.P. Jaehnig, "The ST5000: a high precision star tracker and attitude determination system," in Proceedings of *Space Telescopes and Instrumentation 2001: Infrared and Millimeter*, J.M. Oschmann, M.W.M de Graauw, and H.A. MacEwen, Eds. SPIE Vol **7010**, 70104, 2008.

Phillips, J.D. and R.D. Reasenberg, "Tracking Frequency Laser Distance Gauge," *Review of Scientific Instruments*, **76**, 064501 (2005).

Pollack, S.E., S. Schlamminger, et al. "Temporal Extent of Surface Potentials between Closely Spaced Metals," *Phys. Rev. Lett.* **101**: 071101 (2008).

Reasenberg, R.D., "Aligning a reflection cavity by Anderson's method," Applied Optics, **51(16)** 3137-3144, 2012.

Reasenberg, R.D., J.D. Phillips, M.C. Noecker, High Precision Interferometric Distance Gauge, U.S. Patent 5,412,474, issued 2 May 1995.

R.D. Reasenberg and J. D. Phillips, "A weak equivalence principle test on a suborbital rocket," *Class. and Quant. Grav.*, **27**, 095005, 2010

Reasenberg R.D., Phillips J.D. and Popescu E M, "Test masses for the G-POEM test of the weak equivalence principle," *Class. Quant. Grav.*, 28, 94014, 2011.
.
Robertson, N.A., J R Blackwood, S Buchman, R L Byer, J Camp, D Gill, J Hanson, S Williams and P Zhou, "Kelvin probe measurements: investigations of the patch effect with applications to ST-7 and LISA," *Class. Quant. Grav.*, **23**, 2665, 2006.

Sampas, N.M., Anderson, D.Z., "Stabilization of laser beam alignment to an optical resonator by heterodyne detection of off-axis modes", *Appl. Opt*., **29,** pp 394-403, 1990.

Schlamminger, S., Choi, K.-Y., Wagner, T.A., Gundlach, J.H., Adelberger, E.G, "Test of the Equivalence Principle Using a Rotating Torsion Balance," *Phys. Rev. Lett.* **100**, 041101 (2008).

Schlamminger, S., C.A. Hagedorn, J.H. GundlachSchlamminger, "Indirect evidence for Lévy walks in squeeze film damping," *Physical Review* **D, 81**, 123008 (2010).

Su, Y., B.R. Heckel, et al. "New tests of the universality of free fall." *Physical Review* **D 50**: 3614-3636 (1994).

Sumner, T.J., J. Anderson, J.-P. Blaser, A.M. Cruise, T. Damour, H. Dittus, C.W.F. Everitt, B. Foulon, Y. Jafry, B.J. Kent, N. Lockerbie, F. Loeffler, G. Mann, J. Mester, C. Pegrum, R. Reinhardt, M. Sandford, A. Scheicher, C.C. Speake, R. Torii, S. Theil, P. Touboul, S. Vitale, W. Vodel and P.W. Worden, "STEP (satellite test of the equivalence principle)," *Advances in Space Research*, **39**, 254-258 (2007).

Sumner, T.J., Pendlebury J.M. and Smith K.F. "Conventional magnetic shielding," J. Phys D: Appl. Phys **20**, 1095-1101 (1987).

Sun, K.-X., N. Leindecker, et al. "UV LED Operation Lifetime and Radiation Hardness Qualification for Space Flights." *Journal of Physics: Conference Series* **154**: 012028 (2009).

Thapa, R, J.D. Phillips, E. Rocco and R.D Reasenberg, "Subpicometer length measurement using semiconductor laser tracking frequency gauge," Optics Letters 36(19), 3759-3761, 2011.

Thomas, R.C., J.E. Houston, et al. "The Mechanical Response of Gold Substrates Passivated by Self-Assembling Monolayer Films." *Science* **259,** 1883 (1993).